\documentstyle[prl,aps,epsf,multicol]{revtex}
\begin{document}

\title{Does hardcore interaction change absorbing type critical phenomena?}

\author{Sungchul Kwon$^{1}$, Jysoo Lee$^{2}$ 
and Hyunggyu Park$^{1}$}
\address{$^1$ Department of Physics, Inha University, Inchon 402-751, 
Korea}
\address{$^2$ National Creative Research Initiative Center for 
Neurodynamics \\
and Department of Physics, Korea University, Seoul 136-701, Korea}

\date{\today}
\maketitle
\begin{abstract}
It has been generally believed that hardcore interaction is
irrelevant to absorbing type critical phenomena because
the particle density  is so low near an absorbing
phase transition.
We study the effect of hardcore interaction on the $N$ species 
branching annihilating random walks with two offspring 
and report that hardcore interaction  drastically 
changes the absorbing type critical phenomena in a nontrivial way. 
Through Langevin equation type approach,
we predict analytically the values of the scaling exponents,
$\nu_{\perp} = 2, z = 2, \alpha = 1/2, \beta = 2$ in one dimension 
for all $N > 1$. Direct numerical simulations 
confirm our prediction.  When the diffusion coefficients
for different species are not identical, $\nu_{\perp}$ and $\beta$ 
vary continuously with the ratios between the coefficients.

\end{abstract}

\begin{multicols}{2}


The study of nonequilibrium systems with trapped (absorbing) states
has been
very active in recent years \cite{review}.
Models displaying absorbing phase transitions 
describe a wide range of phenomena, in particular,
epidemic spreading, catalytic chemical reactions, and transport in disordered
media. 
Furthermore, the absorbing transition is
one of the simplest and natural extensions of the well-established
equilibrium phase transition to non-equilibrium
systems, which are still poorly understood.

The concept of universality, which plays a key role in equilibrium
critical phenomena, was shown to be applicable also to
nonequilibrium absorbing transitions.  Critical behavior near an absorbing transition
is determined by properties such as dimensionality and symmetry,
and is not affected by details of the system.  
Finding a new universality class is difficult, and only a few classes of absorbing
transitions are known \cite{review}.

Hardcore interaction between particles or kinks has been believed to
be irrelevant to absorbing type critical phenomena, because the particle density
is so low near an absorbing transition that 
the probability of multiple occupations at a site should be too small to be significant.  
This conventional belief leads to recent successes of field theoretical techniques
using bosonic type operators 
\cite{Doi,Lee,LeeCardy,CardyTauber1,CardyTauber2}. 
However, it is well known that 
hardcore interaction {\em does} changes the asymptotic decay behavior
of the particle density in many multi-species 
diffusion-reaction models near an annihilation fixed point\cite{hb87}. 
Since many absorbing transition models can be
mapped onto diffusion-reaction ones, it may seem natural to 
ask a question whether hardcore constraint changes the absorbing type
universality classes in multi-species models. Despite recent 
efforts using fermionic formulation incorporating hardcore 
interactions \cite{bm99,BOW}, the effect of hardcore interactions 
is barely understood both analytically and numerically.


In this Letter, we study the $N$ species branching annihilating random
walks with two offspring ($N$-BAW(2)), introduced recently by
Cardy and T\"{a}uber \cite{CardyTauber2}.  
The model was solved exactly for
all $N >1$, using renormalization group techniques in bosonic type 
formulation which ignores hardcore interactions. We employ Langevin equation
type approach incorporating hardcore interactions and obtain
analytically the values of critical exponents associated with
the absorbing transition. It turns out that hardcore interaction 
drastically changes the universality class in a nontrivial way
and the critical exponents vary continuously with the ratio of
diffusion constants of different species. Direct numerical simulations
confirm our predictions.


The $N$-BAW(2) model is a classical stochastic system
consisting of $N$ species of particles, $A_i$ $(i=1,\dots, N)$. 
Each particle diffuses on a $d$-dimensional lattice with two competing
dynamic processes: pair annihilation and branching. Pair annihilation is
allowed only between identical particles $(A_i + A_i \rightarrow \emptyset)$.
In the branching process, a particle $A_i$ creates two identical particles 
in its neighborhood $(A_i \rightarrow A_i + 2 A_j)$,
with rate $\sigma$ for $i=j$ and rate $\sigma^\prime / (N-1)$ for
$i \neq j$. 

For $N=1$, this model exhibits an absorbing transition of directed Ising type
($Z_2$ symmetry) at finite branching rate \cite{TT,bALR,KwonPark,Hwangetc}. 
The $N$ species generalization imposes the permutation symmetry $S_N$ between
species. Like in the Potts type generalization of the absorbing transition 
models \cite{Hinrichsen},
this model for $N>1$ is always active except at the annihilation fixed point 
of zero branching rate.


Critical properties near the annihilation fixed point have been explored
exactly by Cardy and T\"{a}uber for $N > 1$ in the framework of
bosonic field theory \cite{CardyTauber2}. 
The upper critical
dimension $d_{c}$ is $2$.  Using a perturbation
expansion, they showed that the branching process associated with
$\sigma$ is irrelevant.
For $\sigma = 0$, it was
found that the models for all $N > 1$  are active for $\sigma^\prime \neq 0$
and their scaling behavior near the annihilation fixed point 
($\sigma^\prime = \sigma^\prime_{c} = 0$ )
forms a new universality class independent of $N$.
For  $d < d_{c}$, the critical behavior is characterized by 
the exponents
\begin{equation}
\nu_{\perp} = 1/d, ~z = 2, ~\alpha = d/2, ~\beta = 1.
\label{eq:exp-nhc}
\end{equation}
Here, the exponents are defined as
\begin{eqnarray}
\xi   & \sim  \Delta^{-\nu_{\perp}}, \ \ \ \ \ \   \tau  & \sim  \xi^{z}, \nonumber \\
\rho (t)  & \sim  t^{-\alpha},  \ \ \ \ \ \ \rho_{s} & \sim  \Delta^{\beta},
\label{eq:exp-def}
\end{eqnarray}
where $\Delta = \sigma^\prime - \sigma^\prime_c$, $\xi$ is the correlation length, $\tau$
the characteristic time, $\rho (t)$ the particle density at time
$t$, and $\rho_{s}$ the steady-state particle density.   

Even in the presence of hardcore interaction, 
the scaling exponents $\alpha$ and $z$ should follow from the 
simple random walk exponents; $z= 2$ and $\alpha= d/z$ for $d<d_c$
\cite{Lee}. Near the annihilation fixed point, elementary scaling theory 
ensures $\beta = \nu_{\perp} z \alpha $, 
which leads to $\beta = \nu_{\perp} d$.
We determine the value of $\nu_{\perp}$ through
Langevin equation type approach.


The particle density can change by branching processes
and pair annihilation processes. 
If we start with a configuration of very low particle density,
the particle density initially grows by branching processes,
$(A_i \rightarrow A_i + 2 A_j)$.
In this growth regime, 
a newly created pair of offspring 
and its parent are far more likely to
annihilate against each other than other particles in the system.
Dynamics of such three particle configurations or ``triplet''
governs the growth behavior of the particle density
and the inter-triplet interactions can be ignored. 
The particle density growth will be finally capped by 
pair annihilations processes of independent particles 
and the system reaches a steady state. 


We focus on the growth regime dominated by triplet dynamics,
from which $\nu_{\perp}$ can be evaluated \cite{CardyTauber2}. 
We only consider the case of $\sigma=0$, where a newly created 
pair is always dissimilar to its parent. The survival probability
$S(t)$ of the triplet of the same species 
decays much faster ($\sim t^{-3/2}$) than that of different species,
so the branching process associated with $\sigma$ is irrelevant.
Near $\sigma^\prime =0$, the time evolution of the particle density of 
$i$-th species, $\rho_i (t)$, is written as
\begin{equation}
{d \rho_{i} \over dt} = 2 {{\sigma^{\prime}} \over {N-1}} 
\sum_{j \ne i} \left[\rho_{j}(t) - 
\int_{0}^{t} L_{ij}(t-t^{\prime})
\rho_{j}(t^{\prime}) dt^{\prime} \right] ,
\label{eq:langevin}
\end{equation}
where $L_{ij}(t-t^\prime)dt$ is the
probability that an $i$-th species pair created by an $j$-th species particle
at time $t^\prime$,
annihilates in an interval between $t$ and $t+dt$.  
The two terms in the right-hand side represent the creation and
annihilation process of a triplet, respectively.
Pair annihilation contribution from independent particles
is ${\cal O}(\rho^{2})$, which is ignored in the growth regime.

The kernel $L_{ij}(t)$ is simply related to the survival probability
$S_{ij}(t)$ of the triplet $(A_j+2A_i)$ as $L_{ij}(t)= -dS_{ij}(t)/dt$.  
To keep the lowest order of $\sigma^\prime$ in 
Eq.~(\ref{eq:langevin}), 
we evaluate $S_{ij}(t)$ at $\sigma^\prime=0$. 
When hardcore interaction
is not present, a pair of $A_i$'s does not see its parent $A_j$, so
annihilate each other freely by random walks. In that case, 
it is well known that $S(t)=S_{ij}(t)$ decays asymptotically 
as $S(t)\sim t^{-\delta}$ with $\delta={1-d/2}$ for $d<2$ and 
becomes finite ($\delta=0$) for $d>2$, 
irrespective of their diffusion constants \cite{bg91}. 
However, with hardcore interaction, the pair annihilation 
process changes significantly due to an effective bias in their diffusive 
behavior, generated by their parent particle $A_j$. 
The motion of $A_i$ near $A_j$ picks up a convective component
with velocity proportional to $t^{-1/2}$, so the 
convective displacement is of the same order of
diffusive displacement $t^{1/2}$. In this case, the competition
between the convection and diffusion becomes nontrivial and the
scaling exponent $\delta$ depends continuously on the 
parameters of the system \cite{s88}.


We calculate the survival probability $S(t)$ of a triplet in one dimension.
With hardcore interaction, $S(t)$ depends crucially on where to 
create two offspring with respect to their parent. When two offspring 
are divided by their parent (static branching) \cite{KwonPark}, 
they have no chance to
meet each other. The survival probability never decays ($\delta=0$).
When two offspring are placed both to the left or both to the right
side of the parent particle with equal probability 
(dynamic branching) \cite{KwonPark},
$S(t)$ decays with a nontrivial scaling exponent. 

Consider three random walkers on a line, labeled as $A$, $B$ and
$C$. $A$ is a parent particle that created two offspring, $B$ and $C$,
to the right side of $A$.
Two offspring $B$ and $C$ are of the same
species, which is different from its parent $A$.  
Hardcore repulsion is present between $A$ and $B$.
$B$ and $C$ annihilate instantaneously upon collision.  
The calculation of $S(t)$ belongs to the class of problems known
as ``capture process'' \cite{bg91,s88,rk99}.  

Let the coordinates of the walkers be $x_{A}, x_{B}$ and $x_{C}$,
and their diffusion coefficients $D_{A}, D_{B}$ and $D_{C}$,
respectively.  In our case, $D_B = D_C$.
It is useful to introduce the scaled
coordinates $y_{i} = x_{i} / \sqrt{D_{i}}$, where $i = A, B, C$.  
Then we can map this triplet system to a single walker
system with isotropic diffusion in three dimensional 
space $(y_A,y_B,y_C)$ \cite{bg91,b88}.
The walker survives inside the wedge bounded by two
planes: a {\em reflecting} plane $P_r$ of
$\sqrt{D_{A}} y_{A} = \sqrt{D_{B}} y_{B}$ 
and an {\em absorbing} plane $P_a$ of
$\sqrt{D_{B}} y_{B} = \sqrt{D_{C}} y_{C}$.

The survival probability $S(t)$ of an isotropic random walker in a $d$ dimensional
cone with absorbing boundary is known \cite{cj59}.  In particular, $S(t)$ in
a wedge with an opening angle $\Theta$ asymptotically decays as $t^{-\pi
/ 2 \Theta}$ \cite{cj59}. In our case, one of the boundary planes, $P_r$,
is not absorbing but reflecting. The probability of the walker at $P_r$
is nonzero and there is no net flux across this plane.
Using this fact, one can easily show that our system should be equivalent 
to the system in a wedge bounded by two absorbing planes 
with twice bigger opening angle. 

We find that the survival probability of the triplet decays 
with the exponent
\begin{equation}
\delta = {\pi \over 4 \Theta} = 
\left[{4 \over \pi} \cos^{-1}\left({1 \over
\sqrt{2(1 + r)}}\right) \right]^{-1},
\label{eq:delta}
\end{equation}
where $\Theta$ is an opening angle of the wedge and 
$r = D_{A}/D_{B}$. The exponent $\delta$ 
monotonically decreases from 1 to 1/2 as 
the diffusivity ratio $r$ varies from 0 to $\infty$.  
At $r=1$ (the same diffusivity for all walkers),
$\delta = 3/4$. 


First, we consider the case that 
diffusion coefficients are identical for all species. 
The $N$ coupled Langevin equations, Eq.~(\ref{eq:langevin}),
can be simplified in terms of the total 
particle density, $\rho(t)=\sum_i \rho_i (t)$, as
\begin{equation}
{d \rho \over dt} = 2 {{\sigma^{\prime}} } 
\rho(t) - 2 {{\sigma^{\prime}}} 
\int_{0}^{t} L(t-t^{\prime})
\rho(t^{\prime}) dt^{\prime} ,
\label{eq:langevin-total}
\end{equation}
where $L(t)=L_{ij}(t)$ is independent of $i$ and $j$.
Taking Laplace transformation, we find
\begin{equation}
s \tilde{\rho}(s) - \rho (0) = 2 \sigma^{\prime} 
(1 - \tilde{L}(s)) \tilde{\rho}(s) =2 \sigma^{\prime} 
s \tilde{S} (s) \tilde{\rho}(s) ,
\end{equation}
where $\tilde{\rho}(s)=\int_0^\infty \rho(t) e^{-st} dt$, and 
similarly $\tilde{L}(s)$ and $\tilde{S}(s)$ are the
Laplace transform of $L(t)$ and $S(t)$, respectively.
With $S(t)\sim t^{-\delta}$, one can show that 
$\tilde{S} (s) \sim s^{\delta -1}$ for $\delta > 0$.

The function $\tilde{\rho}(s)$ has a pole in the positive real axis 
at $s_{o} \sim {\sigma^{\prime}}^{1/(1-\delta)}$.  
When the initial density $\rho(0)$ is small, the density $\rho (t)$ increases
exponentially as $\exp (s_{o} t)$.  
Using the definition of the
characteristic time $\tau$ (Eq.~(\ref{eq:exp-def})), we find
\begin{equation}
\tau = {\sigma^{\prime}}^{-\nu_{\perp} z} = 1 / s_{o} =
{\sigma^{\prime}}^{-1 / (1-\delta )}.
\end{equation}                       

With $\delta=3/4$ for the dynamic branching model, 
we arrive at $\nu_{\perp} z = 1 / (1 - \delta) = 4$.  
Therefore
we predict that
the critical exponents for the 
dynamic branching $N$-BAW(2) model
with hardcore interaction in one dimension are
\begin{equation}
\nu_{\perp} = 2, z = 2, \alpha = 1/2, \beta = 2, 
\label{eq:exp-hc}
\end{equation}
which should be valid for all $N > 1$.  
For the static branching $N$-BAW(2) model, $\delta=0$ and  
$\nu_{\perp} = \beta = 1/2$. Without hardcore interactions, 
branching methods do not matter and $\delta=1/2$, which
leads to Eq.~(\ref{eq:exp-nhc}).


We check the above predictions for the $N$-BAW(2) model by direct
numerical simulations for $N=2, 3$ and 4.  
We start with a pair of particles. 
With probability 
$p$, a randomly chosen particle ($A_i$) creates 
two offspring ($2A_j$) on two nearest neighboring sites 
(dynamic/static branching). The branching probability $p$ 
is distributed as $\gamma p$ for $i=j$ and $(1-\gamma)p/(N-1)$
for $i\neq j$. 
Otherwise, the particle 
hops to a nearest neighboring site.
Two particles of the same species at a site
annihilate instantaneously. In case of models with hardcore
interactions, branching/hopping attempts are rejected
when two particles of different species try to occupy the same
site. Critical probability $p_c=0$ for all models
considered here.

We measure the total particle density $\rho_s$ in the steady state,
averaged over $5\times 10^2 \sim 5\times 10^4$ independent samples
for several values of $\Delta=p-p_c$ ($0.001 \sim 0.05$) and lattice size $L$
($2^5 \sim 2^{11}$). We set $\gamma=1/2$.
Using the finite-size scaling theory \cite{Aukrust}
\begin{equation}
\rho_s (\Delta,L) = L^{-\beta / \nu_{\perp}} F(\Delta
L^{1/\nu_{\perp}}), 
\end{equation}
the value of $\nu_{\perp}$ is determined by
``collapsing'' data of $\rho_s$ with $\beta/\nu_{\perp} = 1$ (Fig.~1). 
Numerical data show that $\nu_\perp$ does not depend on $N$ in all models
as expected.
We find $\nu_\perp = 1.00(5)$ for models without hardcore interactions,
which agrees with the result by Cardy and T\"auber 
\cite{CardyTauber2}.
With hardcore interactions, we find $\nu_\perp = 1.9(1)$ for the 
dynamic branching models 
and $\nu_\perp = 0.50(3)$ for the static branching models, which
confirm our predictions within statistical errors.


When the diffusion coefficients 
are not identical for different species,
$S_{ij} (t)$ decays with the exponent $\delta$
depending on diffusivity ratio $r=D_j / D_i$.
Instead of a single Langevin equation,
we are then forced to deal with the $N$ coupled Langevin
equations.  
The solution of the system of equations is difficult in
general, but the equations become quite simple for $N = 2$.  

Laplace-transformed coupled equations 
for $N =2$ become
\begin{eqnarray}
s \tilde{\rho}_{1}(s) - \rho_{1} (0) & = & 2 \sigma^{\prime} 
s\tilde{S}_{12}(s) \tilde{\rho}_{2}(s) , \nonumber \\ 
s \tilde{\rho}_{2}(s) - \rho_{2} (0) & = & 2 \sigma^{\prime} 
s\tilde{S}_{21}(s) \tilde{\rho}_{1}(s).
\end{eqnarray}
We take $\rho_2 (0)=0$ as an initial condition and
solve the equations for $\tilde{\rho}_1 (s)$: 
\begin{equation}
s \tilde{\rho}_{1}(s) - \rho_{1}(0) = 4
{\sigma^{\prime}}^{2} s \tilde{S}_{12} \tilde{S}_{21}\tilde{\rho}_1 (s).
\end{equation}
Note that $S_{12}(t)$ decays with exponent $\delta(r)$ with 
$r=D_2/D_1$ and $S_{21}(t)$ with $\delta(1/r)$, see 
Eq.~(\ref{eq:delta}).
From the pole position of $\tilde{\rho}_{1}(s)$, we 
arrive at
\begin{equation} 
\nu_{\perp}(r) = {1 \over {2 - \delta (r) - \delta (1/r)}}.
\label{eq:exp-dif}
\end{equation} 
The exponent $\delta(r)$ ranges from $1/2$ to 1, but
$\delta(r)+\delta(1/r)$ varies only slightly with $r$.
It ranges from $3/2$ to 1.5255, so $\nu_\perp (r)$
varies only within a few percent. 
Due to rather large statistical errors ($\sim 10\%$),
we could not confirm numerically the $r$ dependence 
of $\nu_\perp$.
However, it is clear from our derivation that $\nu_\perp$ should vary
continuously with diffusivity ratio.
Although we were not able to obtain a similar expression for
$\nu_{\perp}$ for general $N$, we expect
that $\nu_{\perp}$ varies continuously but only slightly with $r$ 
for all $N > 1$.

In summary, we showed that hardcore
interaction in the $N$-BAW(2) model changes its universality class
in a nontrivial way.
Details of branching methods (static/dynamic branching) and 
also the diffusivity ratios between different species change
drastically the absorbing type critical phenomena.
We find
that, for all $N > 1$, the dynamic branching models with hardcore 
interaction form a new universality class, different from
the models without hardcore interaction. Especially, the scaling
exponents vary continuously with the diffusivity ratios.
The static branching
models with hardcore interaction form yet another new universality class.
Numerical simulations confirm most of our predictions, but
large scale simulations are necessary to measure the 
diffusivity ratio dependence of the scaling exponents.

The present analytic method to study the effect of hardcore
interaction can be applied to a wide range of
multi-species diffusion-reaction models near the annihilation 
fixed point. Our analysis implies that many multi-species 
models with hardcore interaction may exhibit a nontrivial
absorbing phase transition with continuously varying exponents.


We thank the NEST group at Inha University for many useful discussions.  
This work was supported by the Korea Research Foundation for the 21st Century.
J.L. acknowledges support from
Creative Research Initiatives of the Korean Ministry of Science and
Technology.

\end{multicols}

\begin{figure}
\centerline{\epsfxsize=16cm \epsfbox{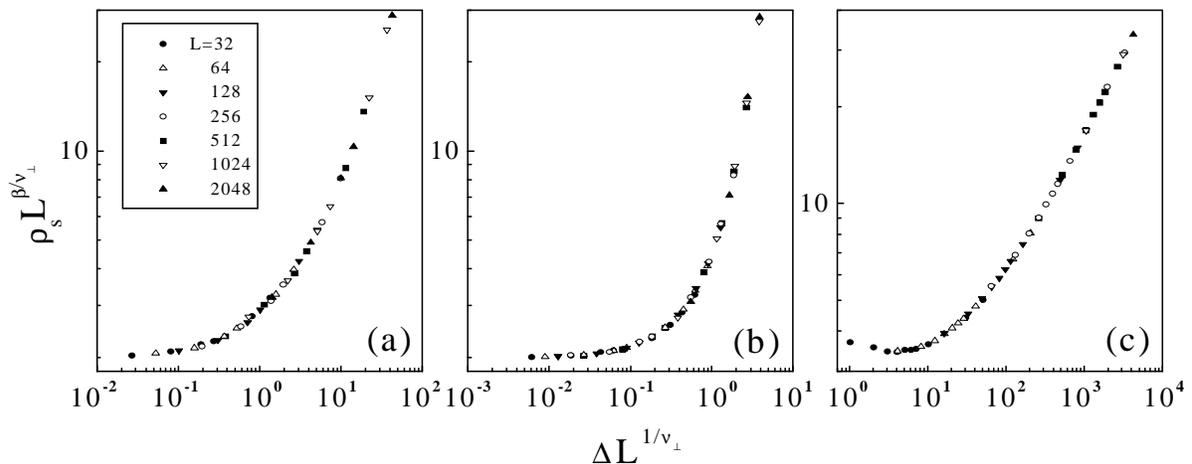}}
\caption{Data collapse of $\rho_s L^{\beta/\nu_\perp}$ 
against $\Delta L^{1/\nu_\perp}$ with $\beta/\nu_\perp=1$ 
for various system size $L=2^5,\ldots,2^{11}$ for $N=2$ BAW(2) models 
(a) without hardcore interaction, 
(b) with hardcore interaction (dynamic branching), 
and (c) with hardcore interaction (static branching).
Best collapses are achieved with (a) $\nu_\perp =1.00(5)$,
(b) $1.9(1)$, and (c) $0.50(3)$, respectively.}
\label{fig-1}
\end{figure}


\begin{references}

\bibitem{review} For a review, see J. Marro and R. Dickman, 
  {\em Nonequilibrium phase transitions in lattice models}
  (Cambridge University Press, Cambridge, 1996);
   H. Hinrichsen, cond-mat/0001070.
\bibitem{Doi} M. Doi, J. Phys. A {\bf 9}, 1479 (1976);
            L. Peliti, J. Phys. (Paris) {\bf 46}, 1469 (1985).
\bibitem{Lee} B. P. Lee, J. Phys. A {\bf 27}, 2633 (1994).
\bibitem{LeeCardy} B. P. Lee and J. L. Cardy, J. Stat. Phys. {\bf 80}, 971 (1995).
\bibitem{CardyTauber1} J. L. Cardy and U. C. T\"auber,
  Phys. Rev. Lett. {\bf 77}, 4780 (1996);
  J. L. Cardy, cond-mat/9607163v2.
\bibitem{CardyTauber2} J. L. Cardy and U. C. T\"auber,  
  J. Stat. Phys. {\bf 90}, 1 (1998).
\bibitem{hb87} S. Havlin and D. ben-Avraham, Adv. in Phys. {\bf 36},
695 (1987); J.-P. Bouchaud and A. Georges, Phys. Rep. {\bf 195},
127 (1990).
\bibitem{bm99} P.-A. Bares and M. Mobilia, Phys. Rev. E {\bf 59}, 1996
(1999).
\bibitem{BOW} V. Brunel, K. Oerding, and F. van Wijland, cond-mat/9911095.
\bibitem{TT} H. Takayasu and A. Yu. Tretyakov, Phys. Rev. Lett. {\bf 68}, 
  3060 (1992).
\bibitem{bALR} D. ben-Avraham, F. Leyvraz, and S. Redner, Phys. Rev. E {\bf 50}, 
  1843 (1994).
\bibitem{KwonPark} S. Kwon and H. Park, Phys. Rev. E {\bf 52}, 5955 (1995).
\bibitem{Hwangetc} M. H. Kim and H. Park, Phys. Rev. Lett {\bf 73}, 2579 (1994);
  W. Hwang, S. Kwon, H. Park, and H. Park, Phys. Rev. E {\bf 57}, 6438 (1998).
\bibitem{Hinrichsen} H. Hinrichsen, Phys. Rev. E {\bf 55}, 219 (1997);
     S. Kwon, J. Lee, and H. Park (unpublished).      
\bibitem{bg91} See, e.g., M. Bramson and D. Griffeath, in {\em Random
  Walks, Brownian Motion, and Interacting Particle Systems}, edited by
  R. Durrett and H. Kesten (Birkhauser, Boston, 1991).
\bibitem{s88} P. Salminen, Adv. Appl. Prob. {\bf 20}, 411 (1988);
           F. Igl\'{o}i, I. Peschel, and L. Turban,
            Adv. Phys. {\bf 42}, 683 (1993).
\bibitem{rk99} P. L. Krapivsky and S. Reder, J. Phys. A {\bf 29}, 5347 (1996);
     cond-mat/9905299.
\bibitem{b88} D. ben-Avraham, J. Chem. Phys. {\bf 88}, 941 (1988);
         M. Fisher and M. Gelfand, J. Stat. Phys. {\bf 53}, 175 (1988).
\bibitem{cj59} H. Carslaw and J. Jaeger, {\em Conduction of Heat in Solids}, 
    (Oxford University Press, Oxford, 1959);
    R. De Blassie, Z. Wahr. verw. Gebiete {\bf 74}, 1 (1987).
\bibitem{Aukrust} T. Aukrust, D. A. Browne, and I. Webman, Phys. Rev. A {\bf 41},
    5294 (1990).

\end{references}
\end{document}